\documentclass[twocolumn,showpacs]{revtex4}

\usepackage{graphicx}

\begin{document}

\title{Nonequilibrium Reweighting on the Driven Diffusive Lattice Gas}

\author{Hwee Kuan Lee and Yutaka Okabe}

\affiliation{Department of Physics, Tokyo Metropolitan University, Hachioji, Tokyo 192-0397, Japan}

\begin{abstract}
The nonequilibrium reweighting technique, which was 
recently developed by the present authors, is used 
for the study of the nonequilibrium steady states. 
The renewed formulation of the nonequlibrium reweighting 
enables us to use the very efficient multi-spin coding. 
We apply the nonequilibrium reweighting to 
the driven diffusive lattice gas model. 
Combining with the dynamical finite-size scaling theory, 
we estimate the critical temperature $T_c$ and the dynamical exponent $z$. 
We also argue that this technique
has an interesting feature that enables explicit calculation of 
derivatives of thermodynamic quantities  without resorting to numerical
differences.
\end{abstract}


\maketitle


Most phenomena occurring in nature are in nonequilibrium states. 
Nonequilibrium systems, such as epidemic~\cite{henkel}, 
vehicular traffic~\cite{chowdhury}, biological network~\cite{berry}, 
and catalysis~\cite{ziff}, have captured a lot of attention.  
Monte Carlo simulation has become a standard tool in scientific computing, 
and advanced simulation methods, such as cluster 
algorithms~\cite{swendsen,hklee} and generalized ensemble 
methods~\cite{berg,lee,oliveira,wang1}, have been developed.  
However, many advanced Monte Carlo methods are not applicable 
to nonequilibrium systems.  Efficient Monte Carlo algorithms 
for nonequilibrium simulation are highly demanded.

Quite recently, the present authors~\cite{hklee1} have developed
a reweighting method for nonequilibrium systems based on the Sequential
Importance Sampling (SIS)~\cite{doucet,liubook}. With nonequilibrium 
reweighting, only simulation at a single temperature is required to obtain
information for a range of temperatures. 
The nonequilibrium reweighting method differs conceptually from 
conventional Monte Carlo methods. In many Monte Carlo methods, a
sequence of micro-states is sampled using the Boltzmann distribution.
One can interpret this as sampling over a ``path" generated by the
associated Monte Carlo updates.
Thermodynamic quantities are then averaged over this path. In 
nonequilibrium reweighting, many paths are first 
sampled with a trial distribution that is not necessarily equal to the 
Boltzmann distribution. Then thermodynamic quantities are calculated 
based on the relative probability between the trial distribution and the
Boltzmann distribution. The relative probability is called ``weights" in
literature, which we shall use hereafter. The advantage of this is that
one could sample many paths at one temperature and 
then calculate required thermodynamic quantities for a range of 
temperatures.

Moreover, Saracco and Albano~\cite{saracco,albano} have proposed 
an effective analysis of nonequilibrium phase transitions, 
in the study of the driven diffusive lattice gas model~\cite{kartz}, 
using a dynamical finite-size scaling theory.  
The behavior of nonequilibrium phase transitions can be extracted 
from short time dynamics~\cite{luo,zheng,okano}. 
If we combine the advantages of dynamical finite-size scaling and 
nonequilibrium reweighting, we can achieve an effective way of 
simulation for nonequilibrium systems.

In this paper, we apply the nonequilibrium reweighting method~\cite{hklee1} 
to the study of the nonequilibrium steady states~\cite{dickman,zia}. 
We illustrate our method on the driven diffusive lattice 
gas model~\cite{kartz}. 
We reformulate the nonequilibrium reweighting method and implement very 
efficient multi-spin coding~\cite{bahnot,roland}.   We also make modifications 
to the dynamical finite-size scaling relation, which was originally proposed 
by Saracco and Albano~\cite{saracco,albano}, so that the advantages of 
reweighting and dynamical finite-size scaling can be combined.


Let us start with explaining the driven diffusive lattice gas model 
proposed by Katz, Lebowitz and Spohn (KLS)~\cite{kartz}. This
system is one of the most well known nonequilibrium models exhibiting 
a nonequilibrium steady state. It was first proposed as
a model for super-ionic conductors, and attained its popularity due to 
its complex collective behavior. It is constructed as a $L_x \times L_y$
square lattice with {\it half-filled} lattice sites having periodic 
boundary conditions. Its Hamiltonian is given by 
\begin{equation}
{\mathcal H} = -4 \sum_{\langle ij, i'j' \rangle} n_{ij} n_{i'j'},
\label{eq:klsham}
\end{equation}
where the summation is over nearest lattice sites. The variable 
$n_{ij} = 1$ when the site is filled and $n_{ij}=0$ otherwise. Attempts for
each particle to jump to an empty nearest neighbor site are given by the
Metropolis rate~\cite{metropolis},
\begin{equation}
T_{\beta,E}(\sigma'|\sigma) = \min \left[ 1, 
\exp ( - \beta (\Delta {\mathcal H} - \epsilon  E ) )\right],
\label{eq:rate}
\end{equation}
where $\sigma$ and $\sigma'$ are the system configurations before and 
after the jump, $\Delta \mathcal H$ represents the change in energy due
to the jump, $E$ is a constant driving force, $\epsilon = -1$, $0$ or 
$1$ depending on whether the jump is against, orthogonal or along the 
direction of the drive, and $\beta = 1/T$ is the inverse temperature 
of the thermal bath.  The $L_y$ direction is taken as 
the direction of the drive. 
The KLS model exhibits an 
order-disorder second order phase transition. In the ordered phase, 
strips of high- and low-density domains are formed along the direction 
of the drive. In the final steady state, the particles are condensed 
into a single strip parallel to the direction of the drive~\cite{hurtado}.
Hence the order parameter can be defined as the density 
profile along the direction of the drive~\cite{saracco}, and moments of
the order parameters are given by
\begin{equation}
\rho^{k} = \frac{1}{(L_x/2)} \sum_{j=1}^{L_x} \left| 
\frac{1}{ L_y}\sum_{i=1}^{L_y} n_{ij} - \frac{1}{2} \right|^{k},
\end{equation}
where $n_{ij} = 0$ or $1$ as defined in Eq.~(\ref{eq:klsham}), and 
$k = 1, 2, 4$ represents the first, second and fourth moments of the order
parameter, respectively.


We briefly review the nonequilibrium reweighting based on SIS, 
and show the implementation on the KLS model.  Define a
path $\vec{x}_t$, a sequence of points in phase space $\sigma_i$ which
were visited during the course of simulation, as
\begin{equation}
\vec{x}_t = (\sigma_1, \sigma_2,\cdots \sigma_t).
\end{equation}
This path can be sampled using the Monte Carlo method at an inverse temperature
$\beta$ and a constant drive $E$. The objective is to calculate 
the appropriate weights for computing 
the thermal average of a quantity $Q$ at another inverse temperature $\beta'$ 
and another drive $E'$,
\begin{equation}
\langle Q(t) \rangle_{\beta',E'} = \sum_{j=1}^n w^j_t Q(\vec{x}^j_t) / 
\sum_{j=1}^n w^j_t,
\end{equation}
where the sum is over all sampled paths indexed by $j$ and $w^j_t$ are 
the weights.  The number of paths is denoted by $n$.  
To calculate the weights, the following steps are 
implemented,
\begin{enumerate}
\item Suppose $\vec{x}^j_t = (\sigma_1^j,\cdots,\sigma_t^j)$ 
up to time $t$ is sampled 
from a simulation at the inverse temperature $\beta$ and drive $E$.
\item To go from $t$, choose a pair of neighboring lattice sites at 
random. If one of the two sites is empty, move the particle to the empty
site with the rate, $T_{\beta,E}({\sigma'}^j|\sigma_t^j)$, 
which is the Kawasaki spin exchange process.  ${\sigma'}^j$
denotes the system configuration after the move.
\item An incremental weight $\delta w^j$ has different 
values according to two possible outcomes,
\begin{enumerate}
\item  If the move is accepted; $\sigma_{t+1}^j = {\sigma'}^j $ and 
$\delta w^j = T_{\beta',E'}({\sigma'}^j|\sigma^j_t) / 
T_{\beta,E}({\sigma'}^j|\sigma^j_t)$.
\item  If the move is rejected; $\sigma_{t+1}^j = \sigma_t^j$ and
$\delta w^j = [1- T_{\beta',E'}({\sigma'}^j|\sigma^j_t)] / 
[1-T_{\beta,E}({\sigma'}^j|\sigma^j_t)] $.
\end{enumerate}
The weights at $t+1$ is given by this incremental weight 
through the relation $w^j_{t+1} = \delta w^j \times w^j_t$ 
with $w^j_1=1$.
\end{enumerate}
For each path $j \in \{ 1, \cdots, n\}$, these steps are repeated until
some predetermined maximum Monte Carlo time is reached.

We make a comment on the technical detail of calculating the weights. 
For case of infinite drive ($E=\infty$), possible values of 
incremental weights $\delta {w_i}$ are,
\begin{equation}
\begin{array}{rcl}
\delta {w_0}&=&1,        \\
\delta {w_1}&=&\exp(-12 (\beta'-\beta)),  \\
\delta {w_2}&=&\exp(-8 (\beta'-\beta)),   \\
\delta {w_3}&=&\exp(-4 (\beta'-\beta))   \\
\delta {w_4}&=& ( 1 - \exp(-12 \beta')) / ( 1 - \exp(-12 \beta))),  \\
\delta {w_5}&=& ( 1 - \exp(-8 \beta') ) / ( 1 - \exp(-8 \beta)) ),  \\
\delta {w_6}&=& ( 1 - \exp(-4 \beta') ) / ( 1 - \exp(-4 \beta)) ). 
\end{array}
\end{equation}
The weights can then be written as a product of incremental weights,
\begin{equation}
w^j_t = (\delta {w_1})^{h_1^j(t)} (\delta {w_2})^{h_2^j(t)} \cdots
	(\delta {w_6})^{h_6^j(t)}.
\label{eq:msw}
\end{equation}
where $h^j_1(t) \cdots h^j_6(t) $ are the number of hits on the 
incremental weights $ \delta w_1 \cdots \delta w_6$ during the course of
simulation from time $1$ to $t$. Note that $\delta w_0$ is irrelevant in
Eq.~(\ref{eq:msw}). Generalization of this counting method to the case 
of finite $E$ is trivial.  Since the calculation of weights has been reduced
to accumulating a histogram, the multi-spin coding technique~\cite{bahnot} 
can be implemented not only for the spin update process but also for 
the calculation of histogram of incremental weights.  For system configuration updates, 
we follow the multi-spin coding technique similar to 
the case of the Kawasaki spin exchange model~\cite{roland}. 
Once the histogram $h^j_1(t) \cdots h^j_6(t)$ is obtained, 
using Eq.~(\ref{eq:msw}) allows us to reweight to a large number 
of temperatures (drives) with negligible extra computational
efforts.  A large increase of efficiency has been obtained 
by a new formulation of the nonequilibrium reweighting. 
The details of the multi-spin coding for the nonequilibrium 
reweighting will be given elsewhere.

For the dynamical finite-size scaling, we use the following equation,
\begin{equation}
\rho^{k} = b^{-\frac{k \beta}{\nu_{\|}}} 
\rho^{* (k)}( b^{-z} \tau, b^{\frac{1}{\nu_{\|}}} \epsilon, b^{-1} L_y,
b^{-\frac{ \nu_{\bot} }{ \nu_{\|}} } L_x, b^{x_0} \rho_0),
\end{equation}
where $k$ is the $k$th moment of the order parameter, $\rho^{*(k)}$ is 
the scaling function for the $k$th moment, $b$ is the spatial rescaling
factor, $\epsilon = (T - T_c)/T_c$, $\beta$ is the critical exponent for
the order parameter (it should not be confused with 
the inverse temperature), $\nu_{\|}$ and $\nu_{\bot}$ are the critical 
exponents for the correlation length parallel and orthogonal to the 
drive, respectively. $z$ is the dynamical exponent and $\tau$ is Monte 
Carlo steps per site. In addition to the original scaling form of 
Saracco and Albano~\cite{saracco,albano}, our scaling form has a term 
$b^{x_0} \rho_0$ to reflect the initial system 
configuration~\cite{okano,zheng}. $x_0$ is an independent exponent and
$\rho_0 \ll 1$ is the order parameter of the initial configuration.
Simulations have to be started with a chosen value of $\rho_0$ for all 
samples. We prepare our initial configuration with $\rho_0=0$ by 
inserting $L_y/2$ particles for each column in the lattice and then 
shuffling each column independently. Letting $b = \tau^{1/z}$, we have 
\begin{equation}
\rho^{k} = \tau^{-\frac{k \beta}{\nu_{\|} z}} 
\rho^{* (k)}( \tau^{\frac{1}{\nu_{\|}z}} \epsilon, 
\tau^{-\frac{1}{z}} L_y,
\tau^{-\frac{\nu_{\bot} }{ \nu_{\|}z} } L_x,
\tau^{\frac{x_0}{z}} \rho_0).
\end{equation}
In the limit of $L_x \rightarrow \infty$ at the critical temperature 
($\epsilon=0$) with $\rho_0=0$, the ratio-of-moments reduces 
to a scaling function with a single argument,
\begin{equation}
\frac{\langle \rho^4 \rangle }{\langle \rho^2 \rangle^2} = 
      g(\tau^{-1/z} L_y)
      \mbox{ with } \rho_0 = 0, \epsilon 
      = 0, L_x \rightarrow \infty.
\label{eq:bscale}
\end{equation}
By plotting the ratio-of-moments versus $\tau L_y^{-z}$ at $T_c$ and 
$\rho_0 = 0$, neglecting corrections to scaling, the curves for 
different system sizes $L_y$ will collapse into a single curve. 
A measure of goodness-of-fit can be 
defined for the ``curve-collapse" as 
\begin{equation}
\eta =  \frac{1}{x_{\max} - x_{\min}} \int_{x_{\min}}^{x_{\max}}
\left|
 g_{L_{y1}} (x) - g_{L_{y2}} (x)
\right| d x,
\label{eq:fit}
\end{equation}
where $g_{L_{y1}}(x) = g( \tau L_{y1}^{-z})$
and   $g_{L_{y2}}(x) = g( \tau L_{y2}^{-z})$.
Our task is to choose $T_c$ and $z$ which minimize $\eta$. 
In using the relation Eq.~(\ref{eq:bscale}), we should check 
that the system size $L_x$ orthogonal to the drive 
is large enough. At this point, we should mention that 
Leung~\cite{leung} has studied the KLS model using finite size
scaling at nonequilibrium steady states. While we focus on
dynamical behaviors, his finite size scaling method was developed
for analysis at stead states.

\begin{figure}
\begin{picture}(0,180)(0,0)
\put(-130,0){\scalebox{.3}{
\includegraphics{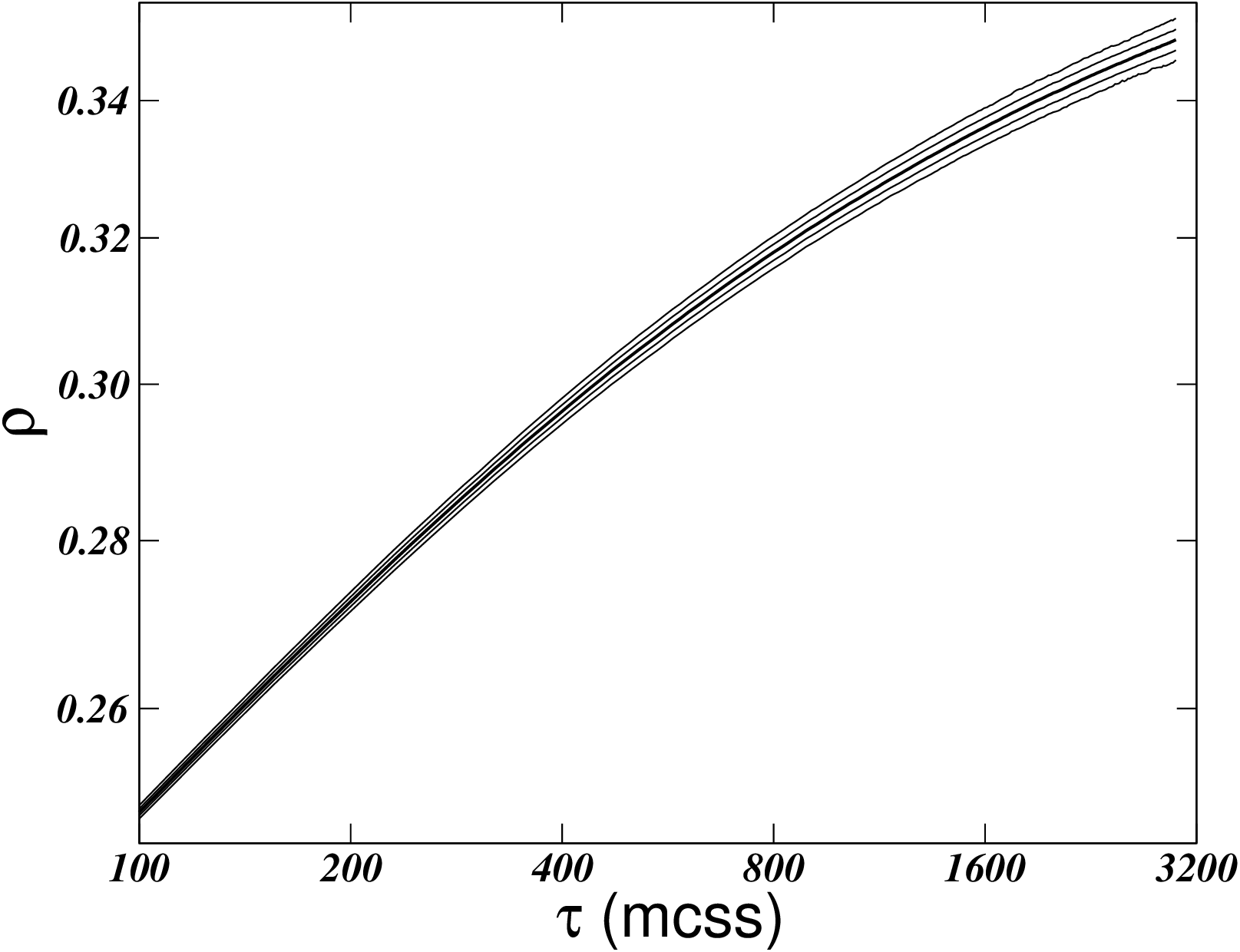}
}}
\end{picture}
\caption{%
Plot of order parameter with infinite drive for $64 \times 32$ lattice 
with actual simulation performed at $T = 3.160$ shown with a bold line.
From top to bottom values of $T$ are 3.150, 3.155, 3.160, 3.165, 3.170.
Averages were taken over $4.096 \times 10^6$ samples.%
}
\label{fig:rhoInf}
\end{figure}

\begin{figure}
\begin{picture}(0,180)(0,0)
\put(-130,0){\scalebox{.3}{
\includegraphics{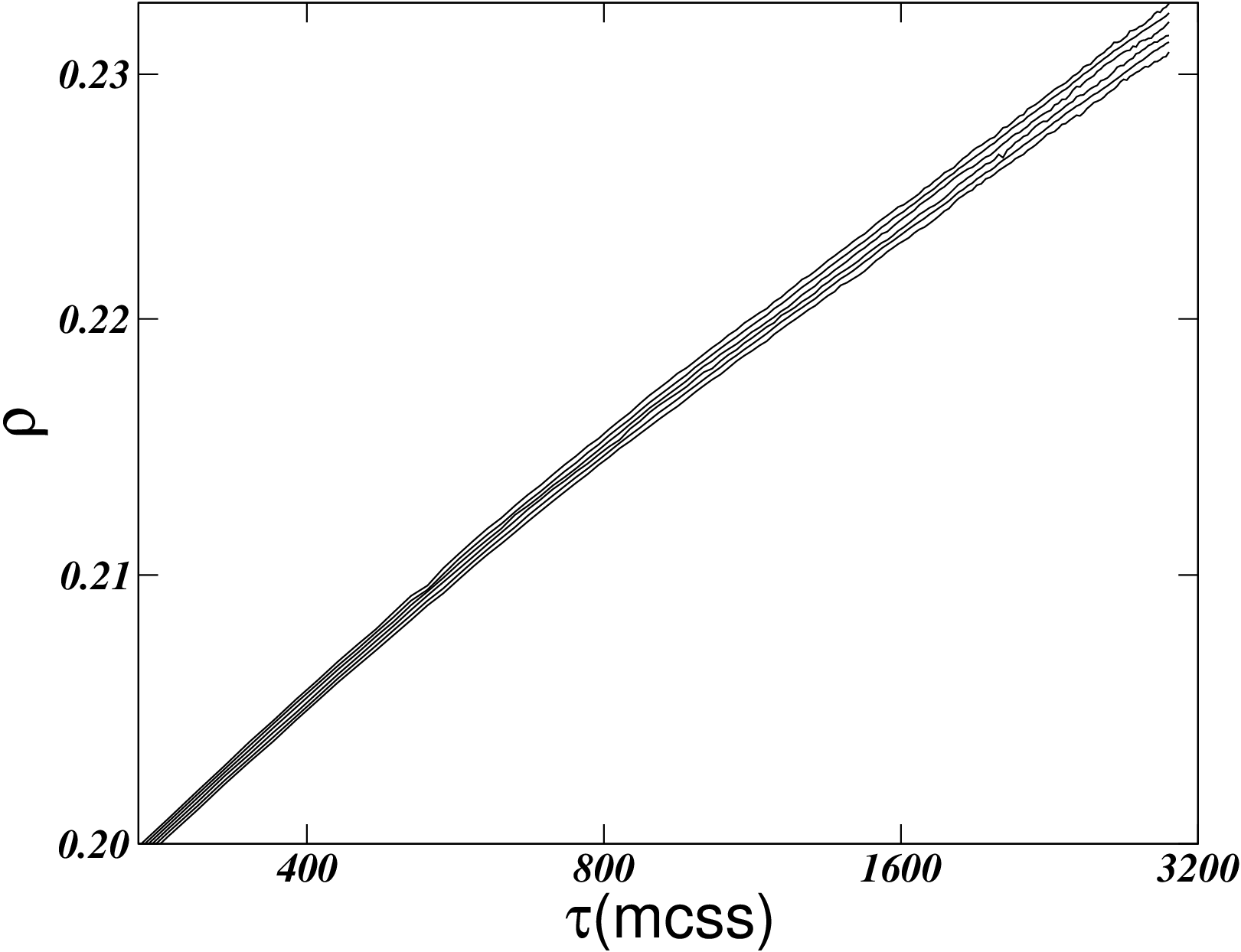}
}}
\end{picture}
\caption{%
Plot of order parameter with finite drive for $32 \times 32$ lattice 
with actual simulation performed at $(T,E) = (2.765, 0.515)$ and 
$(T,E)=(2.780,0.500)$. Reweighted data are combined using weighted mean.
>From top to bottom values of $T$ and $E$ are $(T,E)$ = (2.760,0.520), 
(2.765,0.515), (2.770,0.510), (2.775,0.505), (2.780,0.500), 
(2.785,0.495). Averages were taken over $2.048 \times 10^6$ samples for
each simulation. 
}
\label{fig:rhoE05}
\end{figure}

\begin{figure}
\begin{picture}(0,180)(0,0)
\put(-130,0){\scalebox{.3}{
\includegraphics{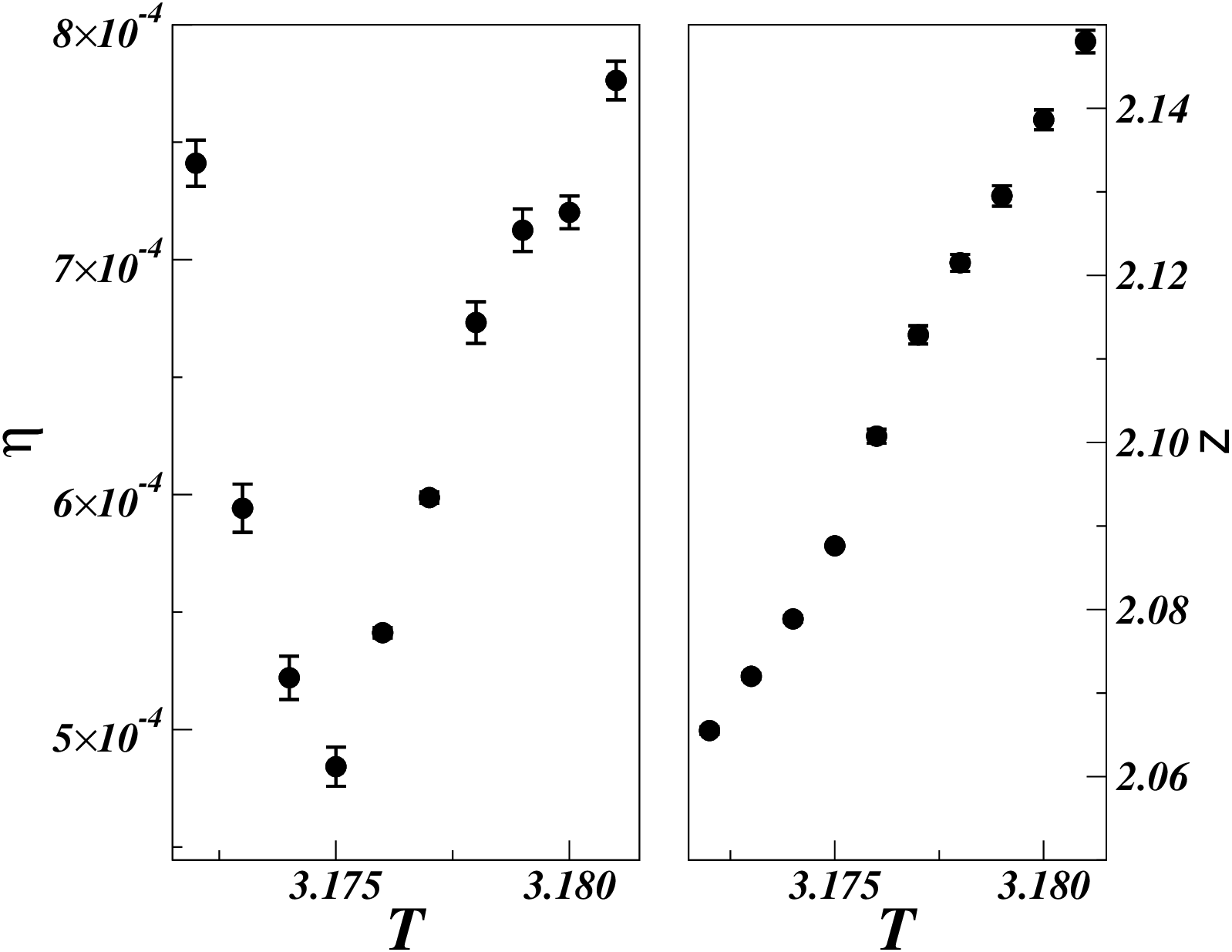}
}}
\end{picture}
\caption{%
Plots showing the goodness-of-fit $\eta$ and corresponding values of
dynamical exponent $z$ for various temperatures. Data are generated by
fitting ratio-of-moments for $L_y=64$ and $L_y=128$ between the range 
$\langle \rho^4 \rangle / \langle \rho^2 \rangle^2 = 
1.2$ and $1.405$. $T_c$ is estimated from the temperature with the 
best fit ($T_c=3.175 \pm 0.002$). Error bars were generated by fitting
several ranges of ratio-of-moments.%
}
\label{fig:fit}
\end{figure}


We now show the results of the Monte Carlo simulation 
for the KLS model. 
We first illustrate the reweighting for the order parameter,
and then show how reweighting can be combined with dynamical finite-size 
scaling (Eq.~(\ref{eq:bscale})) to calculate the critical temperature and
dynamical exponent. 
Figure \ref{fig:rhoInf} shows how data over a range
of temperatures can be extracted from simulations at a single
temperature.  The temporal evolution of the order parameter $\rho$ 
for the infinite drive ($E=\infty$) was investigated 
for $64 \times 32$ lattice. 
Simulations were performed at $T=3.160$, 
and data were reweighted to nearby temperatures, $T=3.150, 3.155, 
3.165, 3.170$ (from top to bottom). Averages were taken over $4.096 
\times 10^6$ samples. We made independent calculations directly at 
$T=3.150$, for example, to check the effectiveness 
of the reweighting. The deviation of the data between the reweighted
ones from $T=3.160$ and the direct ones at $T=3.150$ are 
found to be the same within statistical errors.

We also made simulations for the finite drive ($E \approx 0.5$).
We illustrate the reweighting over both $E$ and $T$. 
We performed two simulations at $(T,E)$ = $(2.765,0.515)$ and 
$(2.780,0.500)$ for $32 \times 32$ lattice.  
The reweighting of the order parameter is made 
by using $\bar{\rho}  = ( \mbox{$\sum_{k=1}^2$} \rho_k / 
\Delta^2_k  )/( \mbox{$\sum_{k=1}^2$} 1 / \Delta^2_k  )$, where 
$\rho_{1,2}$ and $\Delta_{1,2}$ are the order parameter and error 
estimates from the first and second simulations, respectively. 
Figure \ref{fig:rhoE05} shows the temporal evolution of the order
parameter for several temperatures and drives. 
Data was reweighted to several values at $(T,E)$ = $(2.760,0.520)$, 
$(2.770,0.510)$, $(2.775,0.505)$, $(2.785,0.495)$. 
Averages were taken over $2.048 \times 
10^6$ samples for each simulation. Generally, we found that reweighting
is effective when the distributions $P_{\beta,E}(\vec{x}_t^j)$ and 
$P_{\beta',E'}(\vec{x}_t^j)$ have sufficient overlaps. Error bars and 
fluctuations of weights~\cite{liubook} can also be used as quantitative 
measures on the effective range of reweighting.

\begin{figure}
\begin{picture}(0,180)(0,0)
\put(-130,0){\scalebox{.3}{
\includegraphics{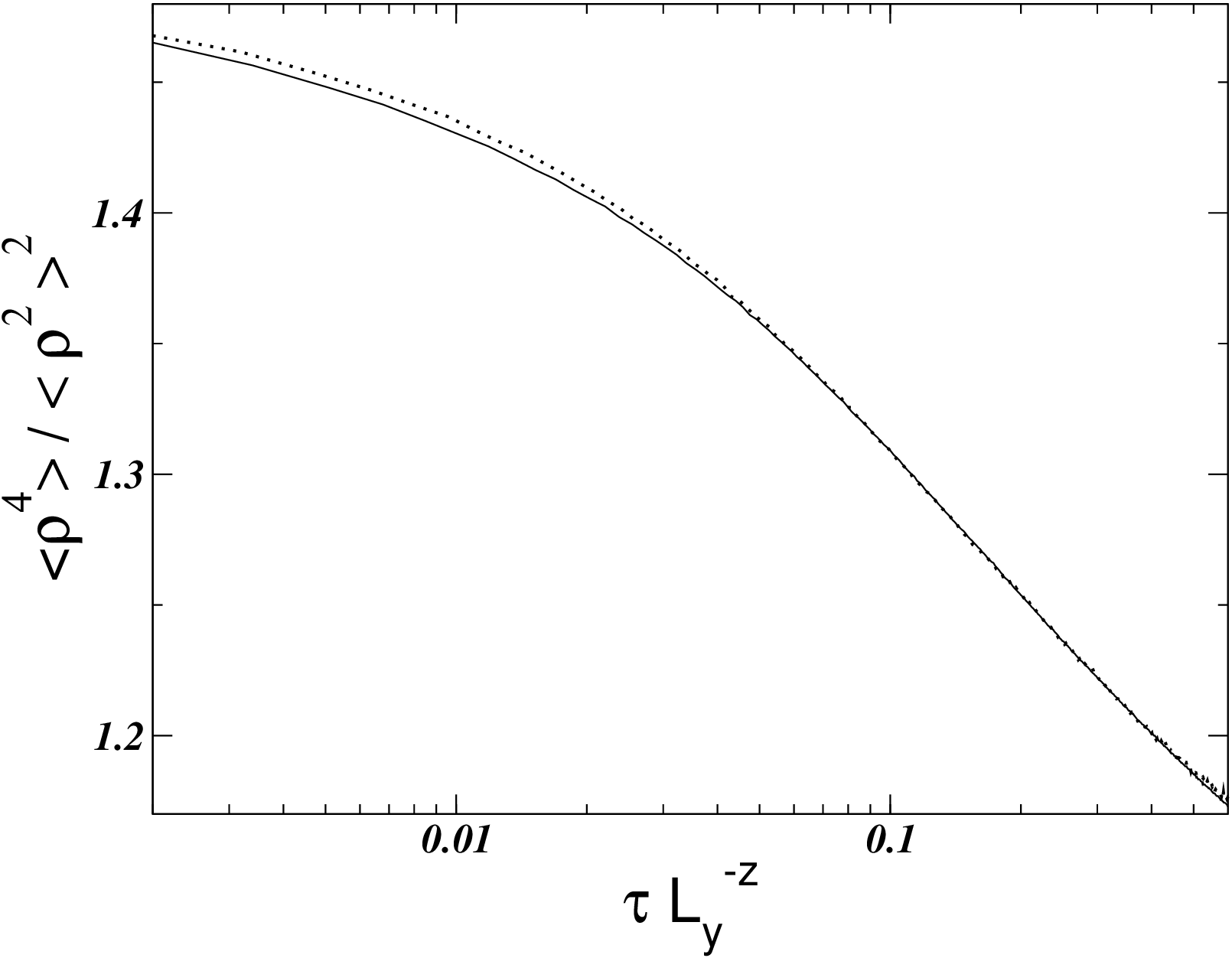}
}}
\end{picture}
\caption{%
Scaling plot of $\langle \rho^4 \rangle/ \langle \rho^2 \rangle^2$
versus $\tau L_y^{-z}$ for $z = 2.09$, $L_y=64$ (solid line) and 
$L_y=128$ (dotted line) at $T=3.175$. Initial system configurations 
were prepared with $\rho_0=0$.%
}
\label{fig:scaling}
\end{figure}

To determine $T_c$, we use the dynamical finite-size scaling 
of the ratio-of-moments 
(Eq.~(\ref{eq:bscale})).  Here we concentrate on the infinite 
drive ($E=\infty$). 
We simulated $64 \times 64$ and $64 \times 128$ lattices, 
and calculated the ratio of the moments, $\langle \rho^4 \rangle/\langle \rho^2 \rangle^2$.  
Before going into the discussion of the fitting, we make 
a comment on the system size $L_x$ whether we can consider 
as $L_x \rightarrow \infty$. 
We performed simulations for both $L_x=64$ and $L_x=128$, 
and confirmed that the ratio of moments for $64 \times L_y$ and 
$128 \times L_y$ coincided with each other to within statistical 
fluctuations. 
Thus, we may regard that $L_x=64$ is large enough. 
Since $\nu_{\|} > \nu_{\bot}$ for the KLS model, 
the correlation length orthogonal to the drive, $\xi_{\bot}$, 
develops slowly; hence $L_x=64$ is large enough to use 
the scaling relation Eq.~(\ref{eq:bscale}).
Now we show the fitting procedure.  
Fitting was performed for several temperatures
near $T_c$, which were reweighted from the data obtained 
at a single temperature, and 
for each temperature we adjusted the value of $z$ such 
that the goodness-of-fit $\eta$, Eq.~(\ref{eq:fit}), becomes minimum. 
Figure \ref{fig:fit} shows $\eta$ 
for several temperatures and the values of $z$ used to calculate $\eta$.
The best fit occurs at $T_c = 3.175 \pm 0.002$; the error bar on $T_c$
is estimated by including all neighboring temperatures where the mean 
values of $\eta$ are within two standard deviations of $\eta$ at 
$T = 3.175$. The value of $z$ within $T=3.175\pm0.002$ is 
$z=2.09\pm 0.01$, and we use this value as our estimate of the dynamical
exponent. Figure \ref{fig:scaling} shows the scaling plot of
$\langle \rho^4 \rangle/ \langle \rho^2 \rangle^2$ as a function of 
$\tau L_y^{-z}$ for $64\times 64$ (solid line) and 
$64 \times 128$ (dotted line) lattice sizes at $T = 3.175$ and $z = 2.09$. 
The curves are almost 
indistinguishable at this scale although some corrections to scaling can
be observed below $\tau L_y^{-z} = 0.02$. To study the corrections to 
scaling, the goodness-of-fit for ratio-of-moments for smaller sizes, 
that is, $64\times 32$ and $64 \times 64$ lattices, 
was also calculated using a similar procedure. 
The best fit occurs at $T = 3.155 \pm 0.005$ with $z = 2.23 \pm 0.03$. 
The estimate for $T_c$ increases with the system size, whereas 
that for $z$ decreases.  Our estimates of $T_c$ and $z$ are 
compatible with the recent estimates for infinite lattice,  
$T_c=3.1980\pm0.0002$~\cite{caracciolo}, $T_c=3.200 
\pm 0.010$~\cite{albano}, $z=2.016\pm0.040$~\cite{albano}. 
A more systematic analysis of the corrections to scaling 
to get a precise estimate of $T_c$ and several critical 
exponents for infinite lattice will be left to a separate 
publication.  Before closing we show the actual procedure 
of the reweighting for each system size.  For $64 \times 32$ lattice, 
$4.096\times 10^6$ samples were used for the simulation
at $T=3.16$. For $64 \times 64$ lattice, $8.19 \times 10^5$ samples 
were used for each simulation at $T = 3.174$ and $3.180$.  
Results were then reweighted to other temperatures and 
combined using weighted mean, $\bar{r}  = ( \mbox{$\sum_{k=1}^2$} r_k / 
\Delta^2_k  )/( \mbox{$\sum_{k=1}^2$} 1 / \Delta^2_k  )$.  Here 
$r_{1,2}$ and $\Delta_{1,2}$ are the ratio-of-moments and error 
estimates from the first and second simulations, respectively.
For $64 \times 128$ lattice size, $1.64 \times 10^5$ samples 
were used for each simulation at $T = 3.174, 3.177$ and $3.180$, 
and reweighted results were combined using the same procedure.

%
%

To summarize, we have studied the use of nonequilibrium reweighting 
based on SIS for the nonequilibrium steady states. 
We have reformulated the nonequilibrium reweighting method, 
which is convenient for the multi-spin coding.  
As a result, a large increase of efficiency has been achieved 
for the performance of simulations. 
We have applied the nonequilibrium reweighting to 
the driven diffusive lattice gas model (the KLS model). 
Combining with dynamical finite-size scaling theory, we have 
estimated $T_c$ and 
the dynamical exponent $z$.

Finally, we make a remark on possible applications. 
The nonequilibrium reweighting method is very general and 
has some very interesting properties.  For example, the 
fluctuation-dissipation theorem does not hold for nonequilibrium systems
and derivatives of thermodynamic quantities had been estimated using 
finite differences~\cite{valles1987}. With reweighting, 
derivatives can be calculated directly by differentiating the 
weights explicitly, that is, 
\begin{equation}
\frac{d \langle Q(t) \rangle_{\beta'}}{d \beta'} = 
\frac{ \sum_{j=1}^n Q(\vec{x}^j_t) \frac{d w^j_t}{d\beta'}  }{
\sum_{j=1}^n w^j_t } - 
\langle Q(t) \rangle_{\beta'} 
\frac{ \sum_{j=1}^n \frac{d w^j_t}{d \beta'}}{ \sum_{j=1}^n w^j_t}.
\end{equation}
Here, $d w^j_t / d \beta'$ can be obtained by differentiating 
Eq.~(\ref{eq:msw}) with respect to $\beta'$. 
We believe that the nonequilibrium reweighting method would have 
several directions for applications.

This work is supported by a Grant-in-Aid for Scientific Research from
the Japan Society for the Promotion of Science. The computation of
this work has been done using computer facilities of the Supercomputer
Center, Institute of Solid State Physics, University of Tokyo.


\end{document}